\documentclass[12pt]{iopart}
\usepackage[pdftex]{graphicx}
\usepackage{amssymb}
\usepackage{subfig}
\usepackage{orcidlink}

\hypersetup{
  colorlinks=true,
  linkcolor=blue,
  urlcolor=black
}

\begin{document}
\pagenumbering{gobble}

\title[Improved Energy-Based Scatter Estimation for PET]{Improved energy-based scatter estimation by incorporating local energy spectra and accelerating the parametric fitting}

\author{ Seyed Amir Zaman Pour$^{1}$ \orcidlink{0000-0002-8816-9358}\,
         Ahmadreza Rezaei$^1$ \orcidlink{0000-0002-9843-8080}, 
         Floris Jansen$^2$ \orcidlink{0000-0002-8437-4960},
         Kristof Baete$^3$ \orcidlink{0000-0003-0113-1590}, 
         Georg Schramm$^1$ \orcidlink{0000-0002-2251-3195},  
         Johan Nuyts$^1$ \orcidlink{0000-0003-1923-6198}
}
\address{ $^1$ University of Leuven, Department of Imaging and Pathology, Nuclear Medicine \& Molecular Imaging, Leuven, Belgium. }
\address{ $^2$ formerly at Department of Engineering, GE HealthCare, Waukesha, WI 53188 USA}
\address{ $^3$ Nuclear Medicine, University Hospitals Leuven, UZ Leuven, Campus Gasthuisberg, Herestraat 49, BE-3000, Leuven, Belgium } 

%%%%%%%%%%%%%%%%%%%%%%%%%%%%%%%%%%%%%%%%%%%%%%%%%%%%%%%%%%%%%%%%%%%%%%%%%%%%%%%%
\begin{abstract}

{\bf Objective:}
Quantitative imaging in positron emission tomography (PET) requires accurate, precise, and efficient scatter correction techniques. Conventional scatter estimation typically relies on single-scatter simulation (SSS) combined with a tail-fitting strategy. However, the accuracy of tail-fitted SSS is limited, for example, by mismatches between the attenuation image and the PET emission data or by the presence of activity outside the field of view (FOV). These shortcomings can be addressed using energy-based scatter estimation (EBSE), as recently proposed by \cite{Efthimiou2022} and \cite{Hamill2024}. The aim of this work is to (1) improve the accuracy of EBSE by accounting for the line-of-response (LOR) dependence of the energy spectrum of unscattered photons, and (2) improve the computational speed of EBSE through better initialization and a more efficient optimization algorithm.

{\bf Approach:} 
The proposed improved EBSE method models the energy probability density function (PDF) of both single and multiple scattered photons, and incorporates a position-dependent (local) energy PDF for unscattered photons. These energy PDFs form the basis of two forward models — a linear nine parameter model and a non-linear five parameter model — used for scatter estimation based on 2D joint energy histograms. The performance of these models were evaluated using GATE Monte Carlo simulations and a NEMA phantom acquisition on a GE SIGNA PET/MR scanner. Furthermore, we assessed the stability of EBSE across the forward models by varying the number of counts in the 2D joint energy histograms via data mashing.

{\bf Main results:} 
EBSE outperformed tail-fitted SSS, particularly in regions near out-of-FOV activity. Our GATE simulations showed that incorporating a local energy PDF for unscattered photons improves off-center regional quantification by approximately 2 percentage points. Additionally, improved initialization combined with the NEGML optimizer reduced the number of required EBSE iterations from 200 to 50, enabling execution on a mashed TOF sinogram in 12 minutes on a modern six-core CPU.

{\bf Significance:}
The proposed method enhances both the accuracy and computational efficiency of EBSE, making it well-suited for clinical applications.

\end{abstract}


%%%%%%%%%%%%%%%%%%%%%%%%%%%%%%%%%%%%%%%%%%%%%%%%%%%%%%%%%%%%%%%%%%%%%%%%%%%%%%%%
\section{Introduction}
Positron emission tomography (PET) is a widely used medical imaging modality that provides insights into functional and metabolic processes within the human body. 
%The primary physical quantity governing PET imaging is the activity concentration of the radiotracer within the subject. 
When positrons emitted from the radiotracer annihilate with electrons, they produce two 511 keV gamma rays emitted back-to-back, which are subsequently detected in coicidence including time-of-flight (TOF).
%Notably, time-of-flight (TOF) PET utilizes the difference in arrival times of gamma rays to more accurately localize the positron annihilation point. 
However, some of these gamma rays undergo Compton scattering within the patient, interacting with electrons in the tissue, reducing their energy and altering their travel direction. This scattering effect introduces image artifacts and degrades image quality when not modeled during iterative image reconstruction, underscoring the need for effective scatter correction techniques \cite{Zaidi2004}.

Single Scatter Simulation (SSS) is a widely used approach for scatter correction in PET \cite{Ollinger1996}. This method utilizes PET emission data and an attenuation image as inputs and follows an iterative, simulation-based process consisting of multiple stages: 1) Initially, emission data is reconstructed with attenuation correction but without scatter correction. 2) A preliminary scatter estimate is then generated using the attenuation map and the non-scatter-corrected activity image. This estimation is based on the Klein-Nishina formula, under the primary assumption that in a pair of photons, only one of the photons Compton scatters once. Subsequently, scaling to the emission data outside the patient (also called tail-fitting) is applied to account for multiple scattering \cite{Watson2005}. 3) The resulting scatter estimate is used to produce a scatter-corrected version of the emission image. 4) The process is then repeated, with the non-scatter-corrected image being replaced by the newly generated scatter-corrected version, and this iteration is continued until the desired result is obtained.

Despite its extensive utilization, the tail-fitted SSS approach has several limitations. 
\begin{enumerate}
\item It relies on an attenuation map, making it susceptible to misalignment between the attenuation map and emission data, particularly due to patient motion between CT and PET acquisitions. 
\item The approach is not applicable in situations where an attenuation map is not available and artifacts in the attenuation image have an impact on the scatter estimate. 
\item For larger objects, the accuracy of scatter estimation is diminished due to limitations in the underlying assumptions. To approximate multiple scattering, the model employs sinogram-based tail fitting; however, this simplification is inherently inaccurate and lacks stability for low-count emission data, which is common in dynamic PET or short scan durations. An alternative method, double-scatter simulation, was introduced by \cite{Watson2018} to eliminate the need for tail fitting, though it increases simulation time. 
\item The SSS approach does not account for scattering from activity outside the field of view (FOV).
\end{enumerate}

Several research groups have investigated energy-based scatter correction methods as an alternative to the tail-fitted SSS approach. In non-energy-based techniques, the energy of incoming photons is used solely to reject low-energy photons using an energy window (e.g. 425-650 keV). In contrast, energy-based scatter correction utilizes the energy of each detected photons to estimate the scatter fraction, addressing many of the limitations associated with the tail-fitted SSS. Notably, this method does not require an attenuation map, eliminating issues related to misalignment between attenuation and emission data. However, energy-based scatter estimation (EBSE) does not require an attenuation map, makes it suitable for joint activity and attenuation estimation, where the initial absence of the attenuation map makes application of SSS rather cumbersome \cite{Rezaei2019}.

In the early stages of research on energy-based scatter correction for PET, technological limitations imposed constraints on the scope of the investigation, with the analysis being performed on only a few energy windows \cite{Bendriem1993, LingxiongShao1994, Grootoonk1996}. 
More recent energy-based methods made use of the detailed energy information provided the listmode data \cite{Popescu2006, Efthimiou2022}. The method of \cite{Hamill2024} fits a nine-parameter forward model to the 2D energy histograms, which accounts for unscattered, single-scattered, and multi-scattered photons in the detected photon pairs.
%However, recent advancements in energy-based methods have been influenced by the contributions of \cite{Popescu2006} and have been further improved and evaluated by \cite{Efthimiou2022, Hamill2024}. These approaches have expanded to incorporate a 2D energy histogram with a forward model that includes nine parameters to account for unscattered, single-scattered, and multi-scattered photons in the detected photon pairs. 
The present study builds on the results of \cite{Hamill2024} and proposes an improved energy-based scatter estimation method as an alternative to the tail-fitted SSS approach in PET imaging. 
Specifically, our work on energy-based scatter estimation includes the following novelties:
\begin{enumerate}
\item An investigation of the dependence of the energy spectrum of unscattered photons (which is needed to model the joint energy
spectrum of all prompt photon pairs) on the incidence angle of the line of response (LOR) and its impact on EBSE.
\item An implementation of a faster maximum likelihood estimator using a better initialization and the NEGML optimization algorithm to achieve substantially faster convergence of model fitting.
\item A forward model for the 2D energy histogram with only five parameters for TOF PET data.
\item Analytically derived models for the energy spectra of the single and multiple scattered photons.
\end{enumerate}
To assess the proposed improvements for EBSE, GATE simulations \cite{Jan2004} and phantom measurements on the GE SIGNA PET/MR were analyzed.

\section{Methods \label{methods}}

\subsection{Theory}
\begin{figure}
    \centering
    \includegraphics[width=0.9\textwidth]{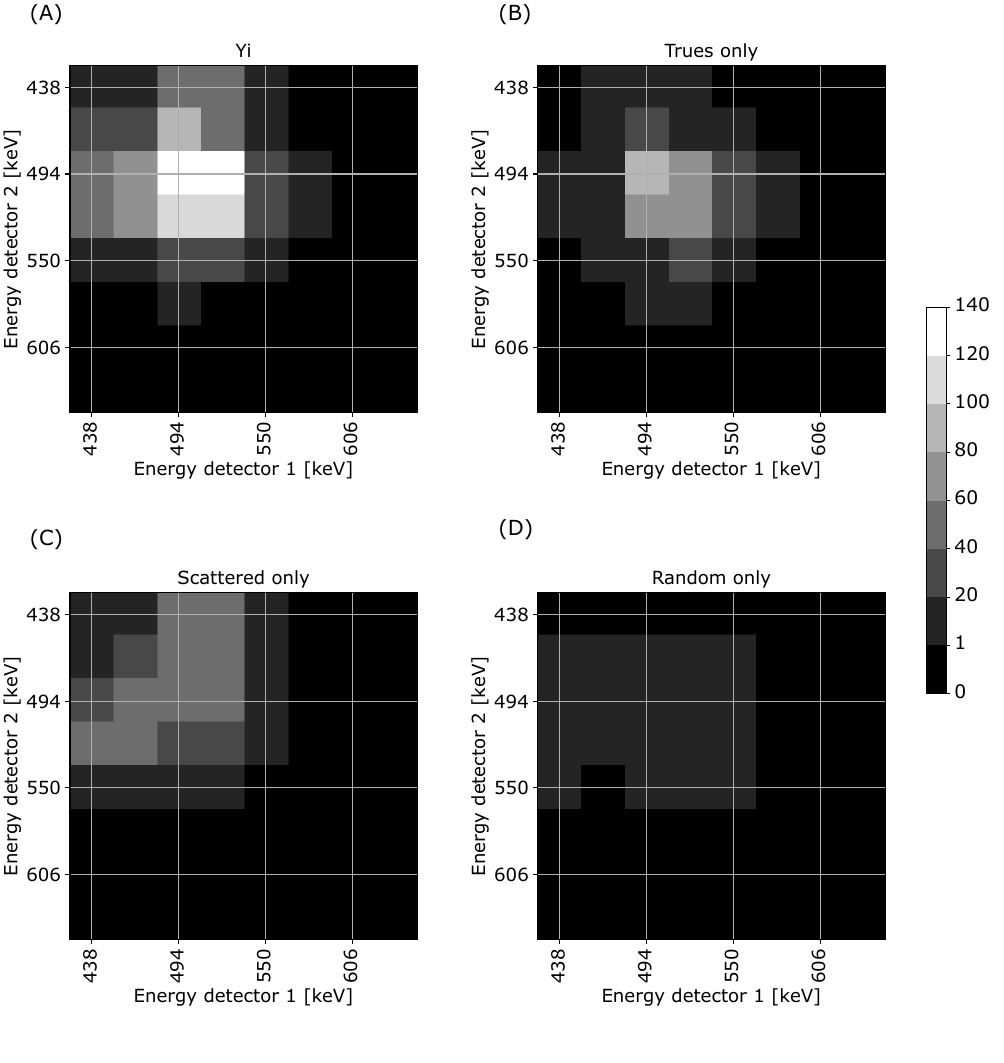}
    \caption{(A) an emission 2D energy histogram of the data collected in the down-sampled sinogram in a single TOF bin. (B) Trues coincidence 2D energy histogram corresponding to (A). (C) Scattered coincidences 2D energy histogram of (A). (D) Random coincidences 2D energy histogram of (A). The x-axis represents the energy of photons detected in detector 1, while the y-axis represents the energy of photons detected in detector 2.}
    \label{fig:figure1}
\end{figure}

The proposed approach involves modeling the measured 2D energy histograms. 
The 2D energy histogram represents the 2D energy distribution of photon pairs
detected in a given TOF bin on a given line of response (LOR).
Figure~\ref{fig:figure1} (A) shows such a 2D energy histogram of GATE simulated data. 
Since this is a simulation, the histogram can be decomposed into unscattered, scattered 
and random coincidences - see Figure~\ref{fig:figure1} (B-D). 
The total prompt coincidence count in the simulation was $7.9 \cdot 10^7$, resulting in an average of approximately 460 counts per 2D energy histogram when using sinogram mashing.
In the following, we use the term ``unscattered'' photon pairs for coincidences where neither photon undergoes Compton interactions within the patient. 

To estimate the fractions of scattered and unscattered photon pairs, we use a model that defines basis functions corresponding to photon pairs with different scatter histories, to model the measured 2D energy histograms. 
One basis function represents the 2D energy distribution of unscattered photon pairs, while all other basis functions correspond to photon pairs with one or more scatter interactions. 
By estimating the weight of each basis function, the relative fractions of unscattered and scattered photon pairs can be determined. 
Since a single photon can have three scatter histories:
"unscattered", "single Compton scatter," or "multi-Compton scatter", 
9 basis functions are used to model the 2D energy distribution \cite{Hamill2024}.
The weights of the basis functions are estimated by optimizing the Poisson log-likelihood function, subject to a non-negativity constraint. 

A full energy-based scatter sinogram can be generated by applying EBSE to the measured 2D joint energy histograms
of all TOF bins along all LORs.
This scatter sinogram can then be used in the maximum likelihood expectation maximization (MLEM) activity reconstruction as the expectation of the scatter contribution.
%This scatter sinogram can then be incorporated into the maximum likelihood expectation maximization (MLEM) activity reconstruction process to model the expectation of scattered coincidences.

The main processing pipeline for EBSE thus consists of the following steps:
\begin{enumerate}
\item Definition of energy probability density function (PDF) of single photons.
\item Construction of 2D forward model basis functions for the 2D joint energy histogram, based on the single photon energy PDFs.
\item Estimation of the weights of these basis functions, which can be used to derive the scatter fraction.
\end{enumerate}
Note that in contrast to SSS or double scatter simulation, EBSE only needs to be applied once before reconstruction directly using
the measured data.

\subsection{Unscattered and single-/multi-Compton scattered single photon PDF} 
To determine the energy spectrum of single unscattered photons, denoted as $P_0$, a prevailing methodology entails the extraction of the single photon energy spectrum from a point source measurement positioned at the center of the scanner. The application of this single PDF to all LORs in energy-based scatter estimation is referred to as EBSE with a ``global unscattered PDF''. 
However, the energy spectrum of single unscattered photons depends on the incidence angle between the detector face and the incoming photon. 
Note that the energy spectrum of single unscattered photons does not conform to a perfect Gaussian distribution and that the probability of energy escape depends on the incidence angle.
This results in a non-Gaussian energy spectrum for single unscattered photons with an elevated lower energy tail that depends on the incidence angle - see Fig.~\ref{fig:figure6}.

\begin{figure}
    \centering
    \includegraphics[width=1\textwidth]{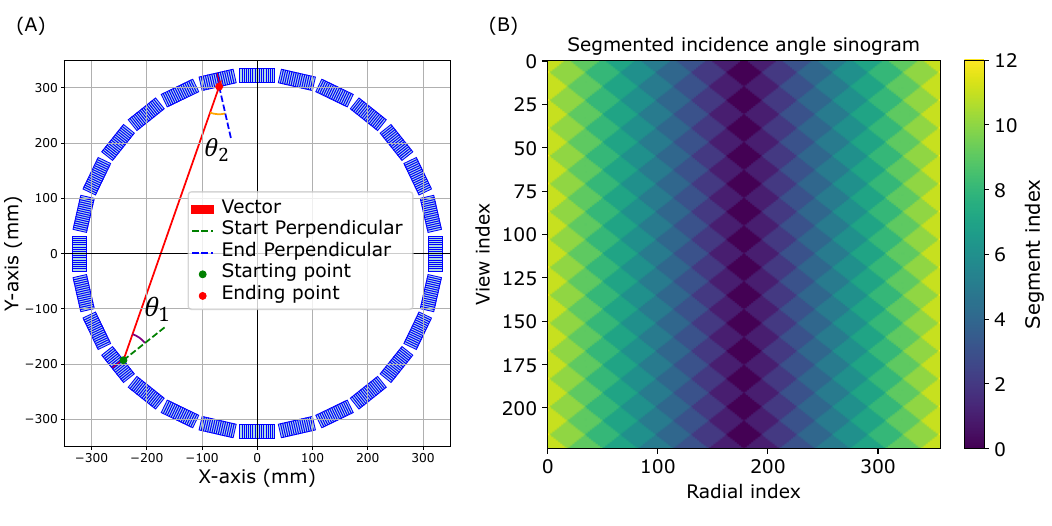}
    \caption{(A) Incidence angle visualization for one LOR in transaxial scanner view.  B) segmented version of the sinogram of incidence angles based on intrinsic patterns.}
    \label{fig:figure2}
\end{figure}

To account for this dependence, we propose to use a local incidence-angle dependent energy spectrum of unscattered single photons
called ``local unscattered PDF''.
The following steps were undertaken to derive the local unscattered PDF.
\begin{enumerate}
\item The incidence angles in 3D space for each LOR were computed. Since the scanner is not a perfect cylinder, the angles between the LOR and the two detectors are not identical. 
Therefore, each LOR has two incidence angles - see Figure~\ref{fig:figure2} (A). % - denoted as $\theta_1$ and $\theta_2$ in Eq.~(\ref{eq:theta}). 
The average incidence angle for each LOR was then determined as $\theta=(\theta_1 + \theta_2) / 2$, 
where $\theta$ represents the incidence angle for that LOR. 
\item The complete sinogram was segmented into 13 distinct regions based on the observed pattern in the computed average incidence angle shown in Figure~\ref{fig:figure2} (B). The average incidence angle ranged from $0^\circ$ to $81.97^\circ$.
\item The energy spectrum for each segment was extracted from a GATE simulation. For real measurements, segmentation was performed using a line-source measurement segmented into three distinct regions. 
This reduction in the number of segments was done to increase the count statistics, thereby achieving a smoother energy spectrum curve for each segment. This down-sampling was performed based on the pattern of incidence angles.
\end{enumerate}

%\begin{equation}
%\theta_{1} = \cos^{-1}\left( \frac{\overrightarrow{LOR} \cdot \overrightarrow{N}_{1}}{|\overrightarrow{N}_{1}| \, |\overrightarrow{LOR}|} \right), \quad
%\theta_{2} = \cos^{-1}\left( \frac{-\overrightarrow{LOR} \cdot \overrightarrow{N}_{2}}{|\overrightarrow{N}_{2}| \, |\overrightarrow{LOR}|} \right)
%\label{eq:theta}
%\end{equation}
%
%Here, $\overrightarrow{LOR}$ represents a vector corresponding to one given LOR in the scanner. $\overrightarrow{N}_{1}$ and $\overrightarrow{N}_{2}$ are normal vectors, perpendicular to the surfaces of detector 1 and detector 2, respectively.

\begin{figure}
    \centering
    \includegraphics[width=1\textwidth]{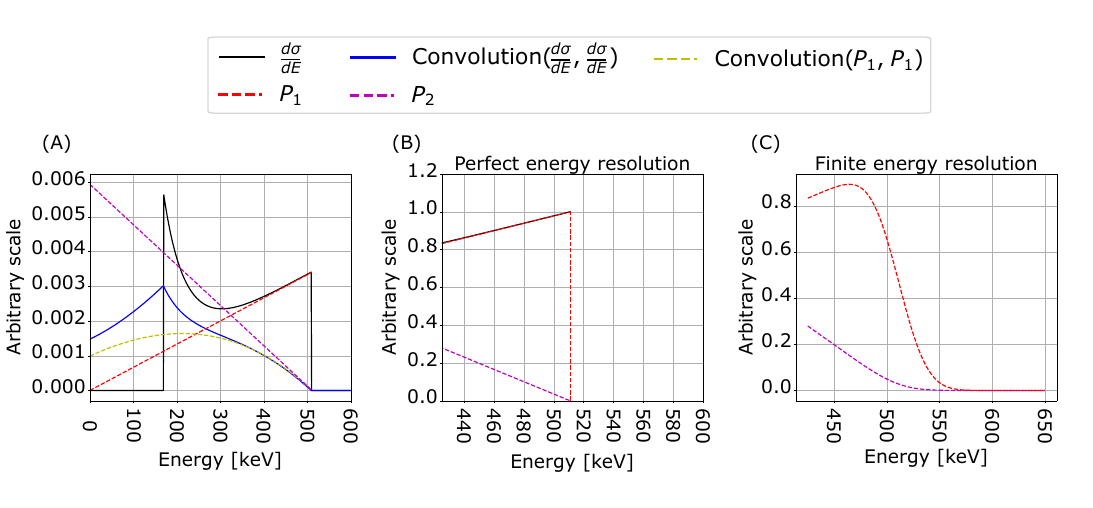}
    \caption{(A) The black curve represents the energy spectrum of single scattering, calculated using the Klein-Nishina formula. The blue curve shows the convolution of the black curve with itself, representing double scattering. The red dashed line corresponds to the model of the single Compton scattered photon above 425 keV ($P_1$). The yellow curve is the analytical convolution of the red dashed line with itself. The purple dashed line represents the first-order approximation of the yellow curve near 511 keV and corresponds to the model of multi-Compton scattered photons above 425 keV ($P_2$).  (B) shows the scattered energy spectrum in the range of the energy window of the SIGNA PET/MR scanner. (C) The energy spectrum of single Compton scattering ($P_1$, red) and multi-Compton scattering ($P_2$, purple) after applying the finite energy resolution model of the scanner as Gaussian smoothing. The y-axis is the arbitrary scale, and the x-axis is energy in keV.}
    \label{fig:figure3}
\end{figure}

Two spectra are employed to characterize the PDF of scattered single photons. 
The first spectrum is an analytical representation of a single Compton scatter PDF, denoted as $P_1$, computed based on the Klein-Nishina equation (black curve) shown in Figure~\ref{fig:figure3} (A) and (B) as a dashed red line. 
The spectrum for multi-Compton scattered single photons, denoted as $P_2$ modeled as an oblique line is shown as a purple dash line. 
The multi-Compton scattered energy spectrum was derived by convolving the Klein-Nishina distribution (black curve) with itself, yielding the double-scattering spectrum (blue curve) shown in Figure~\ref{fig:figure3} (A). 
The analytical method for computing $P_1$ and $P_2$ is provided in Appendix A. 
A Gaussian convolution is utilized to model the scanner's energy resolution, and the full width at half maximum (FWHM) of the kernel was 11.2\%, as determined from point source measurements on the GE SIGNA PET/MR scanner. 
The resulting effective energy spectra of single and multiple scattered single photons are shown in Figure~\ref{fig:figure3} (C). This model is identical to that of \cite{Hamill2024}, where it was justified based on Monte Carlo results.

\subsection{9 and 5 parameter models for the 2D joint energy distribution of photon pairs}
In PET imaging, coincidences are defined by the nearly simultaneous detection of photon pairs. 
The 2D energy histogram, which represents the energy distribution of detected photon pairs, generally requires nine independent parameters for characterization. 
As demonstrated by \cite{Hamill2024}, this model accounts for the inherent complexity of scatter estimation, due to possible correlations between the energies of the two photons of the detected photon pairs. The following definition outlines the nine-parameter forward model:

\begin{eqnarray}
\fl \hat{n}(E_i, E_j) & = &
  a_{0,0} P_0(E_i) P_0(E_j) + a_{0,1} P_0(E_i) P_1(E_j) + a_{0,2} P_0(E_i) P_2(E_j) \nonumber \\
& & + a_{1,0} P_1(E_i) P_0(E_j) + a_{1,1} P_1(E_i) P_1(E_j) + a_{1,2} P_1(E_i) P_2(E_j) \nonumber \\
& & + a_{2,0} P_2(E_i) P_0(E_j) + a_{2,1} P_2(E_i) P_1(E_j) + a_{2,2} P_2(E_i) P_2(E_j)
\label{eq:n_hat}
\end{eqnarray}

The coefficient $a_{0,0}$ is the weight of the $P_0(E_i) P_0(E_j)$ basis function, which represents the number of photon pairs where neither of the photons underwent Compton scattering inside the patient. In contrast, the coefficients $a_{0,1}$ to $a_{2,2}$ collectively represent the number of coincidences where one or both photons experienced single or multiple Compton scattering inside the patient. 
$P_0 (E)$, $P_1 (E)$, $P_2 (E)$ denote the unscattered, single Compton scattered, and multi-Compton scattered single photon energy spectra.
The expectation of the 2D energy histogram of unscattered and scattered photon pairs, denoted as $\hat{n} (E_i,E_j )$, is computed as the weighted sum of  these basis functions. 
Here, $E_i$ and $E_j$ represent the discretized detected energies of the first and second photons, respectively.

We define the expectation of the measured prompt 2D joint energy histogram as 

\begin{eqnarray}
\hat{y}(E_i, E_j) & = & \hat{n}(E_i, E_j) + \hat{r}(E_i, E_j)
\label{eq:y_hat}
\end{eqnarray}

i.e. the sum of $\hat{n} (E_i,E_j )$, along with the expectation for the 2D joint energy spectrum of random coincidences, $\hat{r} (E_i,E_j )$.

To compute $\hat{r} (E_i,E_j)$ for a given TOF emission data bin, the following steps are performed:
\begin{enumerate}
\item The energy spectrum of photons in the delayed window is extracted. Based on this spectrum, a normalized 2D energy histogram is defined under the assumption that the energy distribution of random coincidences is LOR independent. This assumption is validated through the analysis of 2D energy histograms derived from various LORs in GATE-simulated emission data, as described in Section ~\ref{sec:GATESimulation} (results not shown).
\item The normalized 2D energy histogram is then scaled according to the expected number of random coincidences in the given emission data bin, which is provided by the estimated random sinogram.
\end{enumerate}

In cases where the measured energy of a photon in detector 1 is independent of the coincidence photon in detector 2, the 2D energy histogram becomes separable. This means it can be expressed as the product of two 1D energy histograms corresponding to each detector, reducing the number of required parameters to five. This five-parameter model can be particularly applicable to data with good time-of-flight (TOF) resolution, where the improved localization of annihilation events reduces the likelihood of energy correlations.

\begin{figure}
    \centering
    \includegraphics[width=1\textwidth]{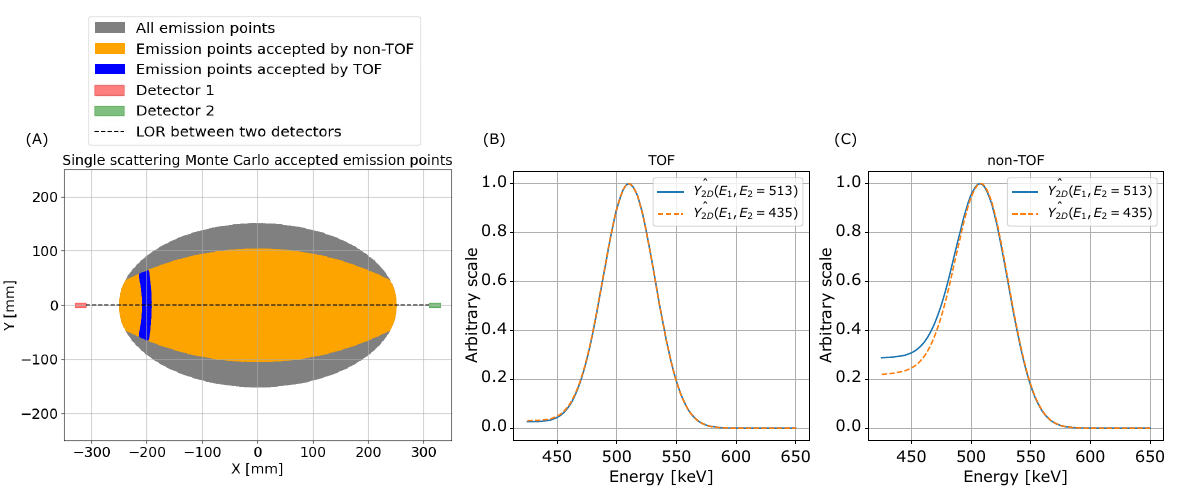}
    \caption{(A) Visualization of emission points that can lead to unscattered or scattered coincidences in a fixed detector pair 
    in the non-TOF (orange) or TOF (blue) case.
    (B) Conditional normalized energy spectra detected in detector 1 when the coincident photon in detector 2 has a fixed energy of 513 keV (blue solid line) and 435~keV (orange dashed line) in TOF (FWHM=380 ps) acquisitions.
    (C) Same as (B), but for non-TOF acquisitions demonstrating that in the non-TOF case the 2D joint energy spectrum is not
    separable due to non independence of the single energy spectra.}
    \label{fig:figure4}
\end{figure}

The correlation between photon pairs arises from uncertainty in their annihilation point along the LOR in non-TOF or low-TOF-resolution emission data. Figure~\ref{fig:figure4} (A) shows the annihilation regions for TOF (blue) and non-TOF (orange) in an elliptical phantom (gray), along with the corresponding detectors in detector 1 and detector 2.

In high-TOF-resolution acquisitions, the region of possible annihilation is relatively narrow. 
This implies that the attenuation experienced by all photons detected in detector 1 (on the left) remains almost constant, and likewise for detector 2 (on the right).  
This is demonstrated by analyzing the energy spectrum in detector 1 when the energy in detector 2 is fixed at 513~keV (predominantly unscattered) and 435~keV (mostly scattered). As shown in Figure~\ref{fig:figure4} (B), the normalized energy spectra for TOF acquisitions remain nearly identical across conditions, indicating statistical independence between the detected photons in detector 1 and detector 2.

In contrast, for non-TOF data, the annihilation point can be situated at almost any point, resulting in more pronounced energy correlations. Figure~\ref{fig:figure4} (C) presents a similar analysis with detector 2 set to two fixed energy bins, revealing differences in the energy spectra observed in detector 1. 
These differences between normalized energy spectra clearly indicate a correlation between the two detected photon energies when TOF
is not available.
In other words, if an unscattered photon is detected in detector 1, the emission point in the non-TOF case is more likely to be closer to the left side due to the effect of attenuation. 
Consequently, the photon detected in detector 2 must have traversed a greater tissue depth, thereby increasing its likelihood of undergoing Compton scattering.

Based on this insight, we propose the following non-linear five-parameter forward model
\begin{eqnarray}
\hat{n}(E_i, E_j) & = & \alpha \Bigl( P_0(E_i) + \beta_1 P_1(E_i) + \beta_2 P_2(E_i) \Bigr) \nonumber \\
                 &   & \quad \times \Bigl( P_0(E_j) + \gamma_1 P_1(E_j) + \gamma_2 P_2(E_j) \Bigr)
\label{eq:n_hat_six_params}
\end{eqnarray}

for the high-TOF resolution case with independent single energy spectra with the coefficients $\alpha$, $\beta_1$, $\beta_2$, $\gamma_1$, and $\gamma_2$. 
$\alpha$ represents the weight of the basis function $P_0 (E_i ) P_0 (E_j )$, corresponding to the number of unscattered (true)
coincidences. 
The nonlinear combination
\begin{equation}
S = \alpha \bigl( \beta_1 + \beta_2 + \gamma_1 + \gamma_2 + \beta_1 \gamma_1 + \beta_1 \gamma_2 + \beta_2 \gamma_1 + \beta_2 \gamma_2 \bigr)
\label{eq:zeta_expression}
\end{equation}
characterizes the number of scattered coincidences.

\subsection{Optimization of model parameters}
The forward model coefficients, as previously described, must be optimized for a given measured 2D joint energy histogram. 
To achieve this, a negative maximum likelihood (NEGML) algorithm \cite{VanSlambrouck2014} approach with a non-negativity constraint is employed, as it is known to accelerate convergence. 
The efficiency of the NEGML algorithm is influenced by the number of parameters updated in each iteration. 
For instance, simultaneously updating all nine parameters per iteration results in a convergence rate similar to that of conventional maximum likelihood expectation maximization (MLEM). 
However, updating the parameters sequentially, one at a time as done in NEGML, leads to a faster convergence rate compared to standard MLEM \cite{VanSlambrouck2014}. 
Additionally, sequential parameter updating is particularly advantageous for the five-parameter forward model, as it is inherently nonlinear. 
Since MLEM and also NEGML are only applicable to linear forward models, the direct application of (NEG)MLEM to the five-parameter model is not feasible. However, updating each coefficient individually while keeping the others fixed ensures that the model remains locally linear at each step, which makes the NEGML algorithm applicable to this optimization problem equation~(\ref{eq:negml}). 
This sequential updating strategy enables the use of (NEG)MLEM for nine and five-parameter forward models. 
The coefficients are estimated using the adapted version of the NEGML approach
\begin{equation}
\Theta_{k}^{(n+1)} = \Theta_{k}^{(n)} + \frac{\sum_{i,j} c_{i,j,k} \frac{y(E_i, E_j) - \hat{y}(E_i, E_j)}{\hat{y}(E_i, E_j)}}{\sum_{i,j} \frac{c_{i,j,k}^{2}}{\hat{y}(E_i, E_j)}} \ ,
\label{eq:negml}
\end{equation}
where $y(E_i,E_j )$ is the measured 2D joint energy histogram, $\Theta_k$ are the coefficients for the nine-/five-parameter models and 
$c_{i,j,k}$ is $\frac{\partial \hat{n}(i,j)}{\partial \Theta_{k}}$.

To accelerate scatter estimation, a weighted least squares estimate was used to initialize the coefficients prior to applying NEGML. 
Marginal spectra $y_1 (E)$ and $y_2 (E)$ are obtained by summing over the other energy dimension in $y(E_i,E_j )$, yielding two 1D prompt spectra corresponding to detector 1 and detector 2. 
The same approach is used to compute $\hat{r}_1 (E)$ and $\hat{r}_2 (E)$ from $\hat{r}(E_i,E_j )$. The observed 1D prompt spectra are modeled as a linear combination of known PDFs representing unscattered, single Compton scattered, and multi-Compton scattered single photons. 
Given the system matrix $A=[P_1 (E), P_2 (E), P_3 (E)]$, where  $P_1 (E), P_2 (E), P_3 (E)$ represent the PDFs, the estimation problem is formulated as:
\begin{equation}
\hat{y}_{l}(E) = A p + \hat{r}_{l}(E), \quad l=1,2
\label{eq:y_l_expression}
\end{equation}

where $p$ is the vector of unknown scaling factors. $l$ represent the index of the first or second dimension. To solve for $p$, we minimize the weighted least squares cost function:
\begin{equation}
L(p) = \bigl( y_{l}(E) - (A p + \hat{r}_{l}(E)) \bigr)^{T} C_{y_{l}}^{-1} \bigl( y_{l} - (A p + \hat{r}_{l}(E)) \bigr)
\label{eq:one_d_init}
\end{equation}

where $C_{y_{l}}$ represents the covariance matrix of the measured spectrum and accounts for statistical noise. 
Since photon detection follows a Poisson process, the covariance matrix is assumed to be diagonal, and an initial approximation is obtained using a Gaussian smoothed (kernel size FWHM = 3 [unit is energy bin width]) estimate of the 1D marginal prompt energy spectra. The parameter estimation is refined iteratively by updating $C_{y_{l}}$ based on the estimated spectrum and solving the weighted least squares. Our observations showed that after two iterations, there is almost no change in the estimated parameters. 
To ensure physical validity, the estimated parameters are constrained to be non-negative. 
This approach provides a computationally efficient and statistically robust initialization.

\subsection{Sinogram downsampling (mashing)}

Sinogram down-sampling was performed by merging detectors in the axial and transaxial directions, a technique known as mashing. However, this process was not applied in the TOF direction, as the scatter is not smooth in the TOF direction, and also, TOF information inherently reduces the correlation between the two photons in a coincidence. 
The primary objective of mashing is to increase the number of counts per 2D joint energy histogram, thereby reducing statistical noise.

The mashing factor (MF) is defined based on the degree of detector merging. For instance, MF = [16,9] indicates that counts from 16 detectors in the transaxial direction and 9 detectors in the axial direction are merged. As a result, the TOF sinogram size of the GE SIGNA PET/MR scanner is reduced from (357, 224, 1981, 27) to (25, 14, 25, 27) radial, angular, planar and TOF bins, respectively.

\subsection{Simulation Experiments} \label{sec:GATESimulation}
Two experiments were conducted to evaluate the proposed energy-based scatter estimation: (1) a GATE simulation and (2) an acquisition of a customized NEMA image quality phantom on the GE SIGNA PET/MR. The GATE Monte Carlo simulations were performed using GATE v9.2, Geant4 11.0.3, and ROOT 6.24/06 on the Ubuntu 20.04.06 LTS operating system.

\begin{figure}
    \centering
    \includegraphics[width=1\textwidth]{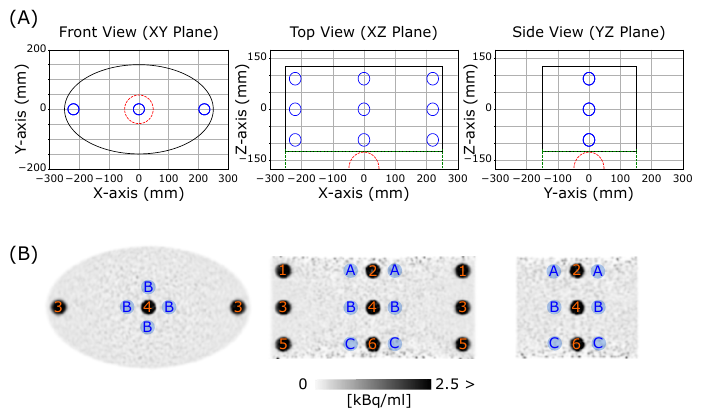}
    \caption{(A): The elliptical phantom geometry used in the GATE simulation is shown in three different plane views. Hot regions within the FOV are depicted in blue, while the red dashed line represents the hot region outside the FOV. The XY plane corresponds to the transaxial view, where the detectors are arranged in a ring, and the Z-axis represents the axial direction. (B) Activity ground truth MLEM reconstruction image of unscattered and non-random coincidences with isotropic Gaussian post-smoothing (FWHM = 2 pixels). The numbers denote the hot regions, while the blue-letter ROI corresponds to the background.}
    \label{fig:figure5}
\end{figure}

A custom elliptical phantom, shown in Figure~\ref{fig:figure5} (A), was defined within the GATE software. The phantom measured 460 mm in the axial direction, with a cross-sectional size of 500 mm x 300 mm. It contained nine spheres, each 37 mm in diameter, arranged in the XZ plane, shown as blue solid lines in Figure~\ref{fig:figure5} (A). Additionally, a larger sphere (96 mm in diameter) was placed outside the FOV, represented by dashed red circle in Figure~\ref{fig:figure5} (A), to simulate activity in the bladder outside the axial FOV.

The phantom was filled with water and a radioactive two-gamma photon emitter at 511 keV. The background activity concentration was 0.25 kBq/ml, while the nine spheres within the FOV were filled with 2.5 kBq/ml. The larger sphere outside the FOV had a concentration of 6.25 kBq/ml. The simulation hyperparameters—including coincidence window, energy resolution, energy window, and coincidence policies—were configured to match the characteristics of the clinical GE SIGNA PET/MR scanner \cite{Levin2016}. The GATE simulation of the elliptical phantom generated $7.9 \cdot 10^7$ prompt coincidences during a 15-minute PET acquisition. 

A key advantage of the GATE simulation was its ability to categorize each detected photon as either scattered or unscattered. This classification was essential, as it enabled the precise reconstruction of unscattered coincidences only, providing a reliable ground 
truth reconstruction, shown in Figure~\ref{fig:figure5} (B). To facilitate a comprehensive comparison between the ground truth and the energy-based scatter-corrected results, the nine sphere boundaries were precisely segmented in the activity images using the geometry map. Due to symmetry between the off-center hot regions, their average was computed for analysis. Additionally, the averages of four spherical segments from three distinct background locations were calculated. The background regions of interest (ROIs) were labeled with blue indices (A, B, and C), while the hot ROIs were denoted by red indices (1–6), as shown in Figure~\ref{fig:figure5} (B). Notably, background ROI C and hot ROIs 5 and 6 were positioned near regions of activity outside the FOV.

To quantify the differences, the relative error was computed for each segment using equation~(\ref{eq:roi_error}).
\begin{equation}
\mathrm{Error of ROI mean} = \frac{ROI_{\mathrm{mean EBS}} - ROI_{\mathrm{mean GT}}}{ROI_{\mathrm{mean GT}}}
\label{eq:roi_error}
\end{equation}

The ROIs are shown in Figure~\ref{fig:figure5} (B). $ROI_{\mathrm{mean EBS}}$ and $ROI_{\mathrm{mean GT}}$ are the mean activity values of a given reconstruction calculated in a given ROI in the ground truth image and energy-based scatter correction image, respectively.

\subsection{NEMA Phantom acquisition}
NEMA phantom measurements were performed on a GE SIGNA PET/MR scanner, with activity positioned outside FOV in the axial direction, as shown in Fig.~\ref{fig:figure11} (C). 
These measurements were reconstructed using both an energy-based scatter correction method and a vendor-implemented, tail-fitted single-scatter simulation estimation approach. The scan was acquired using data from a single-bed position on a GE SIGNA PET/MR scanner operating in research mode. In this configuration, the system was set to provide energy information for single photons in coincidences. The research mode was configured with guidance from the vendor.

The NEMA phantom was filled with 3.67 kBq/ml of $^{18}$F as the background activity. It had a cross-sectional size of 30 cm x 20 cm and contained six spheres with diameters ranging from 10 to 37 mm. The largest and smallest spheres were filled with a 15.18 kBq/ml concentration, while the remaining spheres were kept cold. Additionally, the phantom included a 5 cm cylindrical cold lung-equivalent region.

To introduce outside-FOV scatter contamination, a syringe (volume 100~ml) filled with 73.71 kBq/ml of $^{18}$F was positioned adjacent to the NEMA phantom and fixed within a PMMA phantom. The size of the PMMA phantom corresponded to the typical anatomical dimensions of an adult thorax. The scan acquired a total of $1.4 \cdot 10^9$ prompt counts over a 45-minute duration. To evaluate scatter estimation under realistic count conditions (whole-body static PET scan), the first 2 minutes of the scan were analyzed separately, with prompt counts totaling $7.4 \cdot 10^7$. The estimated scatter distribution from this short scan duration was then scaled to the full scan time.

Additionally, a line-source measurement was performed to derive the local and global energy spectra of unscattered single photons. 
The global unscattered PDF was defined based on the energy spectrum at low incidence angles of the local unscattered PDF. A 30 cm long tube with an activity concentration of 9 kBq/cm was positioned at 10 o'clock in the scanner, 8 cm away from the gantry surface. 
Before extracting the energy spectrum for each incidence angle segment, 
all LORs intersecting the bed were discarded to eliminate photons that scattered in the bed.

For image reconstruction, a listmode  Ordered Subset MLEM (OSEM) algorithm with a voxel matrix of 240 × 240 × 101, an isotropic voxel size
of 2.5 mm, 20 subsets, and 5 iterations for the GATE simulated data, and the NEMA data were used.

\section{Results \label{results}}
\begin{figure}
    \centering
    \includegraphics[width=1\textwidth]{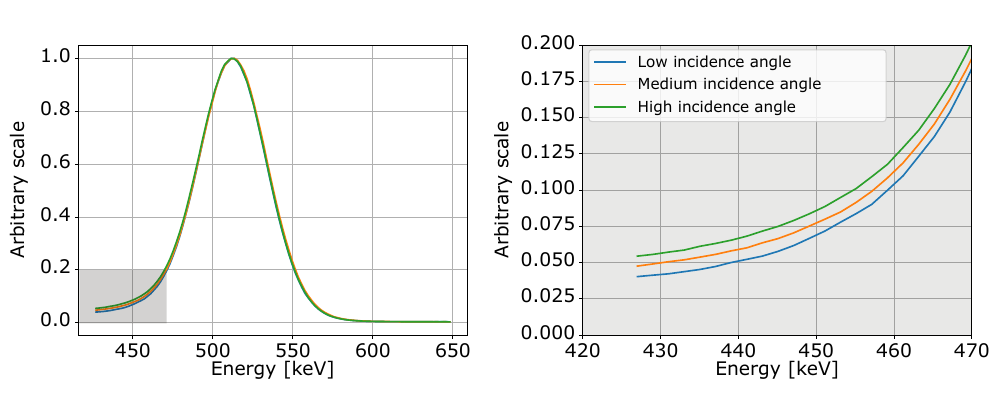}
    \caption{The normalized energy spectrum of unscattered single photons derived from a line-source acquisition on the GE SIGNA PET/MR.
    Note that the right subplot shows a zoomed version of the low energy tails for different incidence angles.}
    \label{fig:figure6}
\end{figure}

Figure~\ref{fig:figure6} shows normalized local energy spectra of unscattered single photons for three sinogram segments, categorized by incidence angle, derived from the line-source measurement. 
As demonstrated, the low energy tail depends on the incidence angle. 
The curve with the highest tail corresponds to the highest-incidence-angle segment (green), 
while the curve with the lowest tail represents the lowest-incidence-angle segment (blue).

\begin{figure}
    \centering
    \includegraphics[width=1\textwidth]{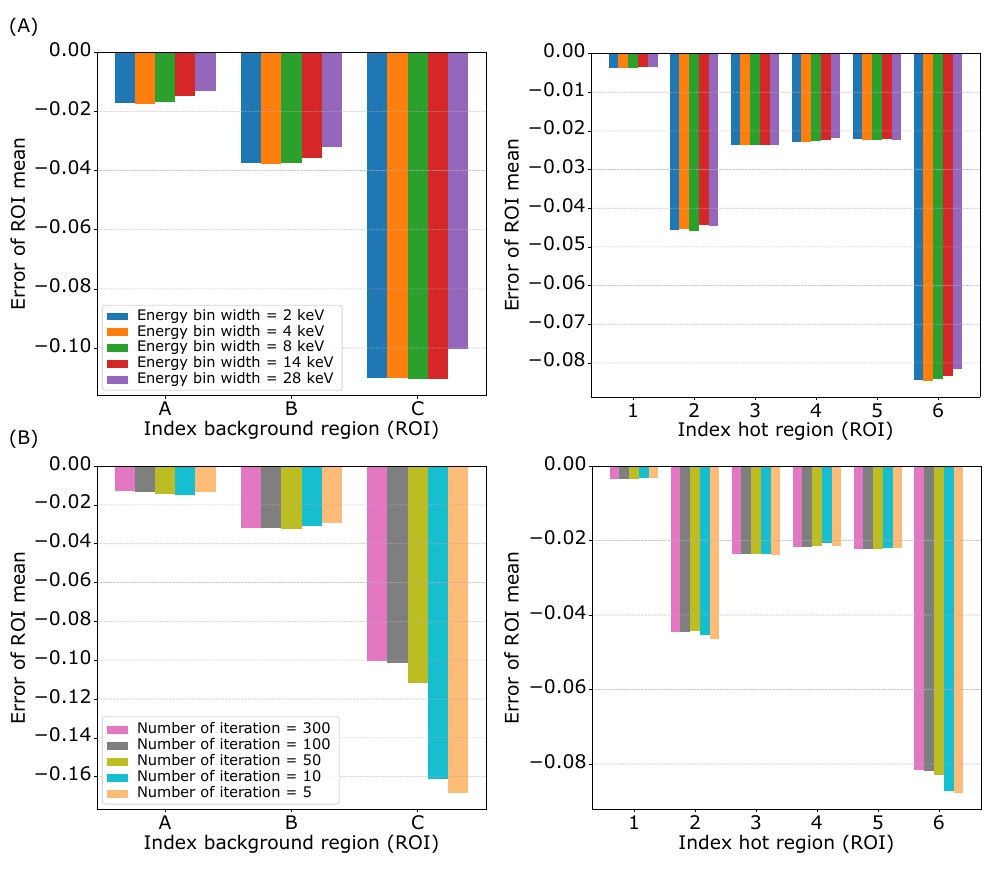}
    \caption{(A) Effect of energy binning on GATE simulated emission data. The Y-axis represents the relative error between energy-based scatter correction reconstruction and reconstruction using only trues coincidences (GT). (B) Selection of the number of iterations for NEGML, ensuring convergence of the estimation process on GATE simulated emission data with initialization coefficients. The Y-axis represents the relative error between energy-based scatter correction reconstruction and reconstruction using only trues coincidences. The X-axis indices correspond to those shown in Figure~\ref{fig:figure5} (B). The nine-parameter method with initialization coefficient is used.}
    \label{fig:figure7}
\end{figure}

Rebinning the 2D energy histogram has a very strong effect on the computational efficiency of the optimization process. To evaluate its impact, five different energy bin widths were tested: 2, 4, 8, 14, and 28 keV. Increasing the bin width progressively reduced the dimensionality of the 2D energy histogram, with its size decreasing from 112x112 to 56x56, 28x28, 16x16, and 8x8, respectively. Among the configurations tested on GATE simulated emission data shown in Figure~\ref{fig:figure7} (A), a bin width of 28 keV was selected for subsequent analyses as it provided an optimal trade-off between computational efficiency and accuracy.

As an iterative estimator, NEGML requires the selection of a hyperparameter, namely, the number of iterations. It is essential to ensure that this value brings the estimation process close to convergence. To determine the minimal number of iterations required for practical convergence, six different iteration numbers were evaluated: 300, 200, 100, 50, 10, and 5. These evaluations were performed after 1D weighted least squares initialization of the coefficients. Our results indicate that the minimum number of iterations required for convergence in GATE simulated emission data is conservatively estimated to be 50 iterations when using our initialization as shown in Figure~\ref{fig:figure7} (B).

%Figure~\ref{fig:figure8} presents the performance comparison between activity reconstructions using energy-based scatter estimated with PDF of global unscattered and PDF of local unscattered. The results indicate that utilizing the PDF of local unscattered yields improved performance compared to the PDF of global unscattered, particularly for hot regions 1, 3, and 5, which are located far from the center.

\begin{figure}
    \centering
    \includegraphics[width=1\textwidth]{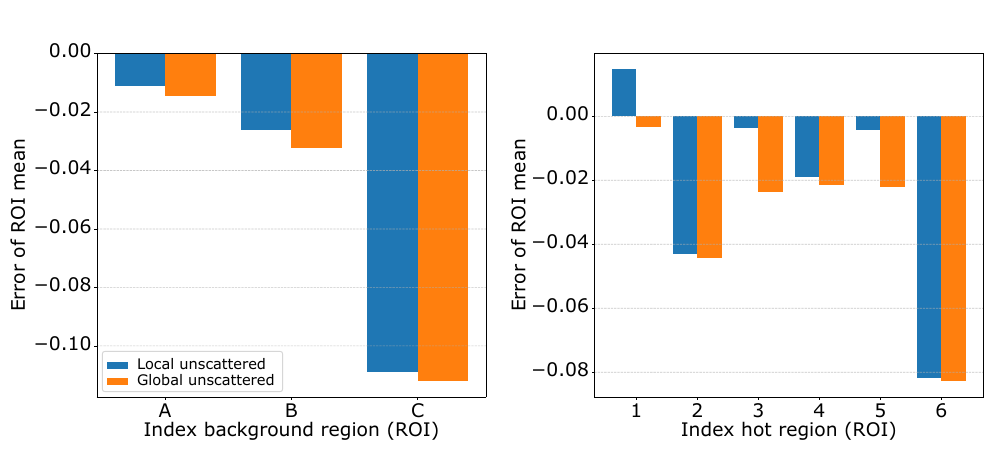}
    \caption{Performance comparison between global and local unscattered PDFs. The Y-axis represents the relative error between energy-based scatter correction and the reconstruction using only trues coincidences. The X-axis indices correspond to those visualized in Figure~\ref{fig:figure5} (B). The nine-parameter method with initialization coefficient is used.}
    \label{fig:figure8}
\end{figure}

Figure~\ref{fig:figure8} presents the performance comparison between activity reconstructions using EBSE with global unscattered and local energy spectra of unscattered single photons. 
The analysis highlights that the effect of using the local unscattered PDF is most evident in off-center regions, specifically in hot regions 1, 3, and 5. The estimated activity distributions are compared with the ground truth activity concentrations used in the GATE simulations to evaluate reconstruction fidelity further. 
This comparison shows a 2 percentage points improvement of EBSE with local unscattered PDF in off-center regions 3 and 5. 
In contrast, minimal differences are observed between the two methods for the central regions 2, 4, and 6.

\begin{figure}
    \centering
    \includegraphics[width=1\textwidth]{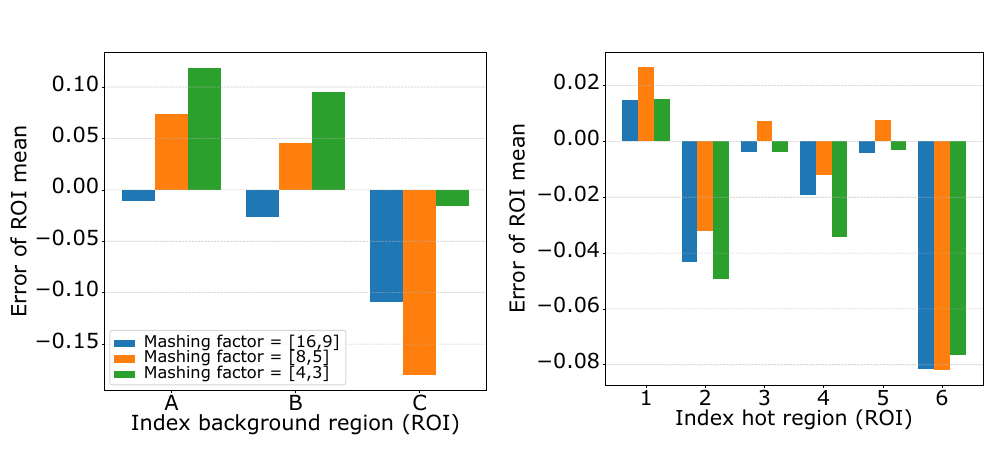}
    \caption{Impact of the mashing factor (MF) on scatter estimation. The left panel represents background regions, while the right panel corresponds to hot spheres. The Y-axis denotes the relative error between energy-based scatter correction and reconstruction using only trues coincidences. The X-axis indices correspond to those shown in Figure~\ref{fig:figure5} (B). The nine-parameter method with initialization coefficient is used.}
    \label{fig:figure9}
\end{figure}

The effect of the mashing factor on energy-based scatter estimation using the nine-parameter model is shown in Figure~\ref{fig:figure9}. As previously outlined, the mashing factor exerts a direct influence on the mean number of counts in the 2D energy histogram. While increasing the mashing factor improves count statistics, an excessively high mashing level can degrade the spatial resolution of the estimated scatter sinogram, thereby affecting scatter estimation accuracy. 
Based on the relative error in both the background and hot regions quantified in the activity reconstructions, all mashing factors exhibit differences smaller than 2\%, indicating their accuracy and reliability.
%Maximum mashing (MF=[16,9]) produced the smallest relative error for the background regions, whereas moderate mashing (MF=[8,5]) or maximum mashing (MF=[16,9]) is accurate for the hot regions quantified in the activity reconstructions.
The scatter fraction (SF) is also investigated for different mashing factors. The ground truth SF for GATE simulated emission data is 42.19\%, while the SF values for each mashing factor are as follows: MF[16,9] = 42.94\%, MF[8,5] = 42.17\%, and MF[4,3] = 43.40\%.

\begin{figure}
    \centering
    \includegraphics[width=1\textwidth]{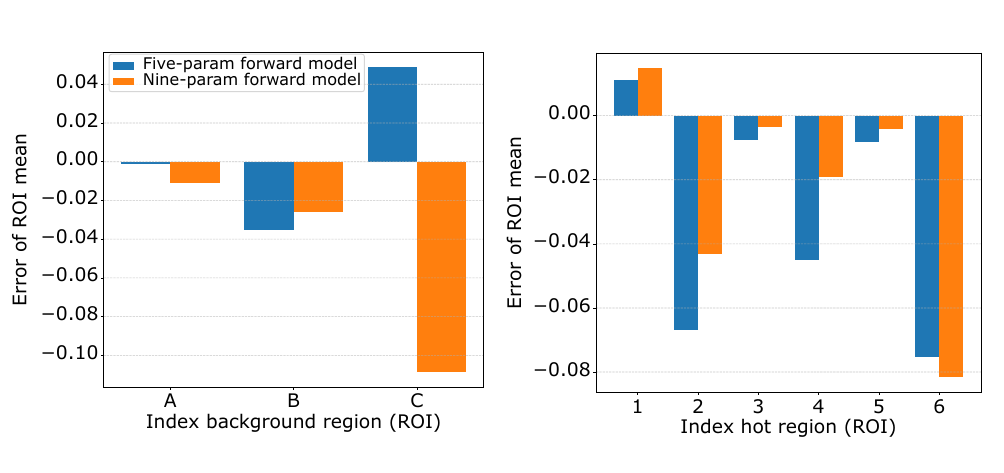}
    \caption{Performance of five- and nine-parameter on the GATE simulated emission data. The Y-axis represents the relative error between energy-based scatter correction reconstruction and reconstruction using only trues coincidences. The X-axis indices correspond to those shown in Figure~\ref{fig:figure5} (B).}
    \label{fig:figure10}
\end{figure}

A comparison between the five- and nine-parameter forward models is shown in Figure~\ref{fig:figure10}. The findings indicate that the nine-parameter model exhibits superior performance in both the hot and background regions. However, the five-parameter model shows better performance in background region index C (near activity outside the FOV).

\begin{figure}
    \centering
    \includegraphics[width=1\textwidth]{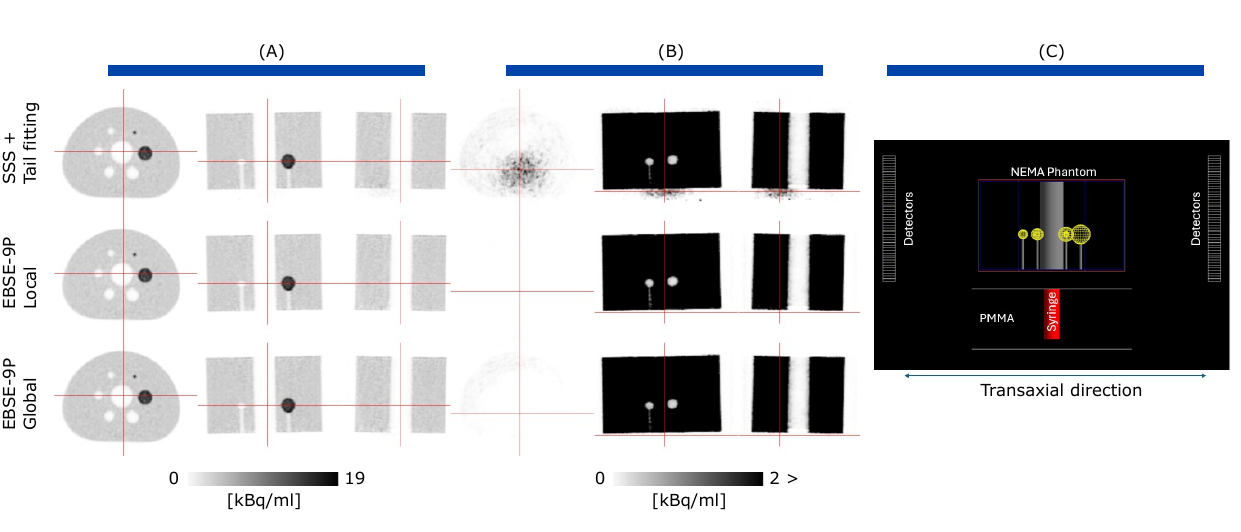}
    \caption{(A) Reconstructed activity images using scatter estimation from two different approaches. (B Same as (A) using a compressed color scale to highlight outside FOV scatter artifacts in the tail-fitted SSS method. (C) The scheme of the NEMA phantom experiment in the GE SIGNA PET/MR scanner (top view). The comparison demonstrates that the energy-based scatter correction method more effectively mitigates artifacts caused by activity outside FOV.}
    \label{fig:figure11}
\end{figure}

\begin{figure}
    \centering
    \includegraphics[width=1\textwidth]{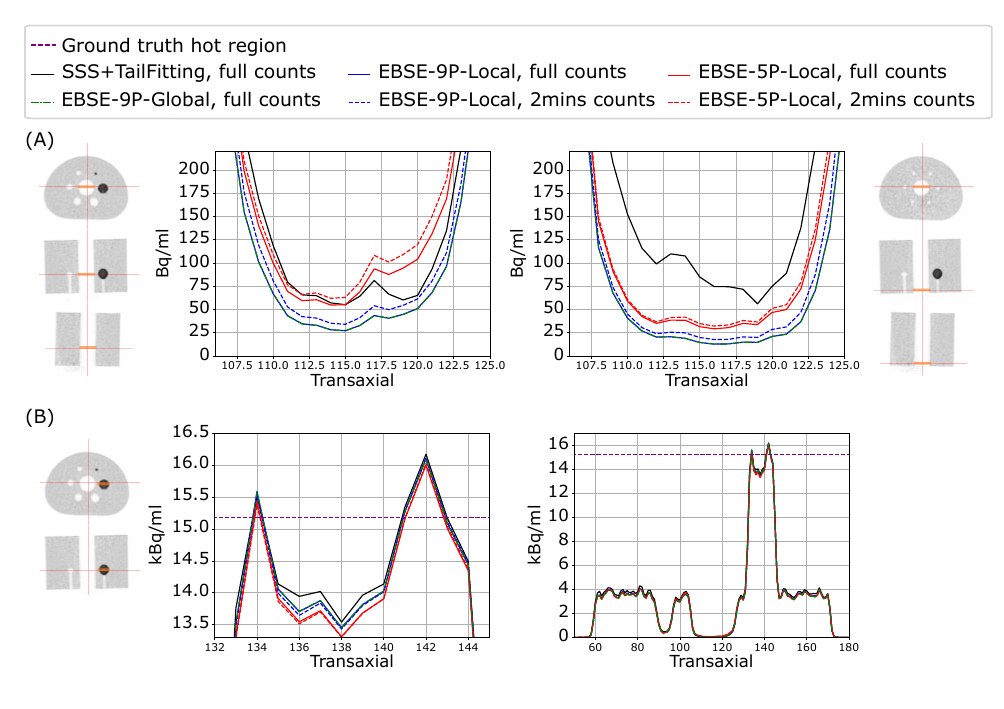}
    \caption{(A) Line profiles through the lung insert of the NEMA phantom reconstruction, comparing different scatter estimation methods across two distinct slices. The left panel corresponds to a slice located far from the activity outside FOV, while the right panel represents a slice positioned closer to the activity outside the FOV. (B) Line profiles through the hot region (37 mm diameter sphere) inserts of the NEMA phantom reconstruction are presented, with different scatter estimations being compared. The right panel corresponds to a profile of the image in a transaxial direction, and the left panel represents the same profile focused on the hot region. The solid line indicates scatter estimation using $1.4 \cdot 10^9$ prompt counts, whereas the dashed line corresponds to estimation using only $7.4~ \cdot 10^7$ prompt counts. The experiment is conducted using a mashing factor of [16,9].}
    \label{fig:figure12}
\end{figure}

The performance of energy-based scatter estimation on real data is evaluated in Figure~\ref{fig:figure11}, which presents a comparison between the tail-fitted SSS, and the energy-based scatter estimation method applied to NEMA phantom measurements. The experiment is conducted using a mashing factor of [16,9]. The results indicate that the energy-based scatter correction approach does not introduce artifacts, even when scattered radiation originates from activity outside the FOV. A quantitative comparison between the two approaches is provided in Figure~\ref{fig:figure12}.

The plots shown in Figure~\ref{fig:figure12} (A), are line profiles intersecting the cavity comparing the nine- and five-parameter forward models using both the local and global unscattered PDFs. The findings indicate that energy-based scatter estimation improves quantification when activity extends beyond the FOV, with the nine-parameter model yielding more accurate results compared to the five-parameter model. There is no difference between using global and local unscattered PDF in the cavity because the lowest incidence angle segment of the local unscattered PDF is used as the global unscattered PDF. Also, the cavity is located at the center of the scanner. Additionally, the nine-parameter energy-based scatter estimation provides slightly better scatter estimation in the cavity compared to the SSS with tail-fitting, even in regions distant from the activity outside the FOV. Figure~\ref{fig:figure12} (B) (right plot) shows the line profile intersecting the hot region, where the horizontal dashed line represents the ground truth activity, derived from the measured activity before injection into the phantom. The results confirm that the energy-based scatter correction method remains reliable for both hot and cold regions, exhibiting minimal influence from activity outside the FOV. 

Furthermore, the difference in scatter estimation between using $7.4 \cdot 10^7$ prompt counts from a 2-minute scan (dashed curved line) and $1.4 \cdot 10^9$ prompt counts from a 45-minute scan (solid curved line) in Figure~\ref{fig:figure12} is minimal, indicating the robustness of the method across different count levels. 

\section{Discussion \label{Discussion}}
Building on the work of \cite{Hamill2024}, this study explores and validates extensions that further enhance the performance of the energy-based scatter estimation method. 
Our findings indicate that the energy spectrum of unscattered photons is dependent on the incidence angle. To assess the relevance of this observation, we investigated how this angular-dependency influences the energy-based scatter estimation. Our approach incorporates a position-dependent (local) energy spectrum that accounts for incidence angle variations. Moreover, we present an analytical derivation for the scattered photons energy spectra, which confirms the model proposed by \cite{Hamill2024} based on results from Monte Carlo simulations.
%In contrast to them, the approach presented here utilizes a local energy spectrum of single unscattered photons that varies with the incidence angle, rather than an energy spectrum of global unscattered photons.
%Furthermore, the energy distribution of scattered photons is modeled analytically rather than being numerically extracted, providing a more adaptable framework for scatter estimation.

Utilizing the NEGML optimization algorithm with initialization significantly reduces the required number of iterations, thereby accelerating the estimation process. The results in GATE simulated emission data from the hot regions indicate that the difference between using 300 iterations and as few as 50 iterations is negligible. However, using fewer than 50 iterations is not recommended in the background regions. Notably, a difference 
 of $~6\%$ was observed between the results obtained with 300 iterations and those with 10 or 5 iterations. 
Findings reported by \cite{Hamill2024} indicate that standard MLEM without advanced initialization required 200 iterations for the nine-parameter forward model.
%Incorporating findings from \cite{Hamill2024}, standard MLEM without advanced initialization required 200 iterations for the nine-parameter forward model.

The second approach to reduce computation time involved rebinning the 2D energy histogram. Theoretically, a reasonable energy bin width can be approximated as half of the energy resolution. Based on our $^{18}$F point source measurement, the GE SIGNA PET/MR scanner demonstrated an energy resolution of 11.2\%. Our results from GATE emission data indicated that the difference between using a fine energy bin width of 2 keV and a coarse energy bin width of 28 keV was less than 1\% in both hot regions and backgrounds. This energy bin-width increase reduced the 2D energy histogram size from 112 x 112 to 8 x 8 which significantly reduced in computation time, making the estimator approximately 32 times faster. 
In our optimized setting the down-sampled scatter sinogram (mashing=[16,9]) estimation process took 12 minutes using six CPU cores
(AMD EPYC 7773X, 64-Core Processor, 2.2 GHz).

The impact of the incidence angle dependence is negligible in the center and changes the activity by only 2\% in the off-center region. This indicates that for most applications, and brain imaging in particular, using a single, low incidence angle energy spectrum is sufficient.
%The impact of using a local versus a global energy spectrum of unscattered single photons is most pronounced in off-center hot regions (ROIs 1, 3, and 5) in the GATE simulated data. 
This occurs because, at low incidence angles, the global energy spectrum of unscattered single photons closely resembles the local spectrum of unscattered photons, making their distinction less significant in central regions compared to off-center. 
The observed similarity between the energy spectrum of a point source and the lowest-incidence-angle segment in the GATE simulated emission data justified using the lowest-incidence-angle segment as the global energy spectrum of unscattered single photons for NEMA phantom scatter estimation. 
However, we think the distinction between different angle classes in the energy spectrum of unscattered photons becomes increasingly important for large axial-FOV scanners, as they are likely to detect a greater fraction of events at higher angles of incidence.

The findings of this study indicate that, in a comparison between the nine-parameter and five-parameter energy-based scatter estimation forward models, the nine-parameter model demonstrates superior performance under realistic and high-count conditions. This improved performance can be attributed to the additional degrees of freedom, which enable a more precise estimation of scattered events. Furthermore, it was hypothesized that the nine-parameter forward model could be effectively replaced by the five-parameter if 
high-resolution TOF information was available. 
However, for the five-parameter model to achieve comparable accuracy, higher TOF resolution may be required. 
The computation times for the nine- and five-parameter models were similar due to the non-linearity of the five-parameter model which makes the optimization slightly more computation intensive, which offsets the gain from estimating fewer parameters.
%A notable drawback is the inherent non-linearity of the five-parameter model, which complicates the implementation of optimization algorithms for coefficient estimation.

Adjusting the mashing factor influences two critical parameters: the average count per 2D energy histogram and the spatial resolution of scatter estimations. Increasing the mashing factor reduces statistical noise and accelerates computation. 
Results from GATE simulated emission data suggest that the relative error changes of the reconstructed activity caused by changes to the mashing are small (<2\%).
%Results from GATE simulated emission data suggest that the optimal mashing factor varies depending on the region of interest. A mashing factor of [16,9] is optimal in background regions, while a lower factor of [8,5] is recommended for hot regions. Increasing the mashing factor extends the area over which scatter estimation is performed, potentially averaging activity contributions from both background and hot regions, which may reduce local accuracy. Conversely, reducing the mashing factor to [4,3] degrades performance due to lower count statistics, leading to a slight bias, as indicated by the scatter fraction analysis.

Overall, the recommended strategy for energy-based scatter estimation is to employ the nine-parameter model with appropriate initialization, a high level of mashing, and the use of a global energy spectrum for single unscattered photons. This strategy is applicable under the condition that a 2\% error (off-centered) is considered negligible in studies involving large objects.
%the recommended strategy for energy-based scatter estimation is to use the nine-parameter model with proper initialization, high mashing level, and incorporation of a global energy spectrum of single unscattered photons if in your study 2\% is negligible for large objects.

\section{Conclusion \label{Conclusion}}
This study demonstrates that energy-based scatter estimation outperforms SSS with tail fitting when activity is present outside the FOV, while performing at least as well as vendor-implemented SSS with tail fitting in all other regions. Additionally, this method remains applicable in cases where an attenuation map is unavailable, such as in joint attenuation and activity reconstruction \cite{Rezaei2019}. By incorporating the NEGML estimator with improved initialization, the estimation process is significantly accelerated, reducing the required number of iterations per 2D energy histogram. Further speed improvements are achieved by down-sampling the 2D energy histogram, maintaining computational efficiency with minimal impact on accuracy. Accounting for incidence angle in the unscattered energy spectrum improves scatter estimation, though the effect is minimal in central FOV regions, suggesting that a global energy spectrum may be sufficient for small objects like in brain imaging. Lastly, while both the five-parameter and nine-parameter models demonstrated stability in simulations and phantom measurements, the nine-parameter model exhibited superior accuracy. However, further investigation is needed to determine the minimum count requirements for each method, particularly as the five-parameter model is only applicable to TOF emission data.

%%%%%%%%%%%%%%%%%%%%%%%%%%%%%%%%%%%%%%%%%%%%%%%%%%%%%%%%%%%%%%%%%%%%%%%%%%%%%%%%

%%%%%%%%%%%%%%%%%%%%%%%%%%%%%%%%%%%%%%%%
\section*{Acknowledgments}

We would like to express our sincere thanks to everyone who contributed to the success of this study. In particular, we are grateful to Timothy Deller for providing insightful discussions throughout this research. This work was supported in part by the Research Foundation—Flanders (FWO) under grant G062220N. The Department of Imaging and Pathology at KU Leuven received funding from GE HealthCare for this research project.

%%%%%%%%%%%%%%%%%%%%%%%%%%%%%%%%%%%%%%%%
\appendix
\section*{Appendix: Analytical model for the scatter energy spectrum \label{appendix A}}
\setcounter{equation}{0}
\renewcommand{\theequation}{A.\arabic{equation}}
\renewcommand{\thesubsection}{\arabic{subsection}}

The (single photon) scattered energy spectrum is modeled as a combination of single scattered and multiple scattered.

\subsection{Single Compton scattered energy spectrum $(P_1)$}
The fraction of the energy that is transferred from the incoming photon (at 511 keV) to the scattered photon is given by Compton’s equation:
\begin{equation}
f(E_0, \theta) = \frac{E}{E_0} = \frac{1}{1 + \frac{E_0}{m_e c^2} (1 - \cos \theta)}
\label{eq:energy_scatter}
\end{equation}

where $E_0$ is the energy of the incoming photon, $E$ the energy of the scattered photon, $\theta$ the scattering angle and $m_e c^2$ = 511 keV. This equation can be reorganized to compute the angle from the energies of the incoming and scattered photons:
\begin{equation}
\theta(E_0, E) = \arccos \left( 1 - \frac{m_e c^2}{E_0} \left( \frac{E_0}{E} - 1 \right) \right)
\label{eq:theta_energy}
\end{equation}
The single scattered energy spectrum we wish to compute corresponds to the derivative of the scatter cross-section with respect to the energy and can be computed as:
\begin{eqnarray}
\mathrm{SingleScatteredSpectrum}(E_0, E) & = & \frac{d\sigma}{dE}(E_0, E) \nonumber \\
& = & \frac{d\sigma}{d\Omega}(E_0, \theta) \cdot \frac{d\Omega(\theta)}{d\theta} \cdot \left| \frac{d\theta(E_0, E)}{dE} \right|
\label{eq:single_scattered_spectrum}
\end{eqnarray}

The differential cross section for single scattering with angle $\theta$ is given by the Klein-Nishina formula:
\begin{equation}
\frac{d\sigma}{d\Omega}(E_0, \theta) = K(E_0, \theta) = \frac{1}{2} r_e f^{2}(E_0, \theta) \left( f(E_0, \theta) + \frac{1}{f(E_0, \theta)} - \sin^{2}\theta \right)
\label{eq:compton_diff_cross_section}
\end{equation}

where $r_e$ is the classical electron radius, which can be ignored here, because we are only interested in the shape of the energy spectrum.

The solid angle $d\Omega(\theta)$ covered by all scattered directions with angle $\theta$ is the area of a circular strip with radius $\sin \theta$ and width $d\theta$. Consequently
\begin{equation}
\frac{d\Omega(\theta)}{d\theta} = 2 \pi \sin \theta
\label{eq:solid_angle_derivative}
\end{equation}

The derivative of the angle $\theta (E_0,E)$ with respect to $E$ equals
\begin{eqnarray}
\frac{d\theta(E_0, E)}{dE} & = & \frac{d}{dE} \arccos \left( 1 - \frac{m_e c^{2}}{E_0} \left( \frac{E_0}{E} - 1 \right) \right) \nonumber \\
& = & - \frac{\frac{m_e c^{2}}{E^{2}}}{\sqrt{1 - \left( 1 + \frac{m_e c^{2}}{E_0} - \frac{m_e c^{2}}{E} \right)^{2}}}
\label{eq:dtheta_dE}
\end{eqnarray}

Inserting these three derivatives into equation~(\ref{eq:single_scattered_spectrum}) produces the energy spectrum of the single scattered photons in PET, which is shown as the black curve in Figure~\ref{fig:figure3}:
\begin{eqnarray}
\mathrm{EnergySpectrum}(E_0, E) & = & \pi f^{2}(E_0, \theta) 
\left( f(E_0, \theta) + \frac{1}{f(E_0, \theta)} - \sin^{2} \theta \right) \nonumber \\
& & \cdot \frac{m_e c^{2}}{E^{2}} \cdot \sin \theta \nonumber \\
& & \times \frac{1}{\sqrt{1 - \left(1 + \frac{m_e c^{2}}{E_0} 
- \frac{m_e c^{2}}{E} \right)^{2}}}
\label{eq:energy_spectrum}
\end{eqnarray}

For application to PET, we set $E_0$=511 keV. To simplify the model for application in an energy window near 511 keV, we use a linear approximation, i.e. the line tangent to $\mathrm{EnergySpectrum}(E_0,E)$ at $E$=$E_0$ = 511 keV. Somewhat lengthy calculations (without approximations) produce
\begin{equation}
\lim_{E \to E_0} \frac{d}{dE}\,\mathrm{SingleScatteredSpectrum}(E_0, E)
= \frac{1}{E_0}
\label{eq:limit_singlescattered}
\end{equation}

Consequently, the linear approximation is a line from the origin (0, 0) to the rightmost point (511, EnergySpectrum(511,510)) of the energy spectrum, which is shown as the red dashed line in Figure~\ref{fig:figure3}. 

\subsection{Double-Compton scattered energy spectrum $(P_2)$}
In the energy window around 511 keV, the multiple scatters are dominated by double-Compton scattered photons. An approximation of the double-scattered energy spectrum can be obtained by convolving the single-scattered spectrum equation~(\ref{eq:energy_spectrum}) with itself. The computation is further simplified by convolving the linear approximation with itself. Writing this linear approximation as $aE$, with $a$ the slope and $E$ the energy, one obtains for the double-scattered energy spectrum:

\begin{eqnarray}
\mathrm{DoubleScatteredSpectrum}(E_0, E) & \simeq & a^2 \Bigl( 
  E_0^2 (E_0 - E) 
  - E_0 (E_0 - E)^2 \nonumber \\
& & \quad + \frac{(E_0 - E)^3}{6} 
\Bigr)
\label{eq:double_scattered_spectrum}
\end{eqnarray}

As a first-order approximation near $E_0$ only the first-order term is kept: $\mathrm{DoubleScatteredSpectrum}(E_0, E)
\simeq a^2 \Bigl(E_0^2 \,(E_0 - E)\Bigr)$. This linear approximation is shown in Figure~\ref{fig:figure3} (A) as the purple dashed line. The figure also shows plots of the (numerical) convolution of the single scattered spectrum with itself (blue), and the (analytical) convolution of the linearized single scattered spectrum with itself (yellow dash-line). In this work, a weighted sum of the linearized single and double energy scatter spectra, convolved with the Gaussian energy resolution, was used Figure~\ref{fig:figure3} (C).

%%%%%%%%%%%%%%%%%%%%%%%%%%%%%%%%%%%%%%%%%%%%%%%%%%%%%%%%%%%%%%%%%%%%%%%%%%%%%%%%
\section*{References}
%\nocite{*}
%\bibliographystyle{apalike}
%\bibliography{EnergyBasedScatter}

\end{document}